# Towards 100,000-pixel microcalorimeter arrays using multi-absorber transition-edge sensors

S.J. Smith[1,2,*], J.S. Adams[1,2], S.R. Bandler[1], S. Beaumont[1,2], J.A. Chervenak[1], A.M. Datesman[1,3], F.M. Finkbeiner[1,4], R. Hummatov[1,2], R.L. Kelly[1], C.A. Kilbourne[1], A.R. Miniussi[1,2], F.S. Porter[1], J.E. Sadleir[1], K. Sakai[1,2], N.A. Wakeham[1,2], E.J. Wassell[1,3], M.C. Witthoeft[1,5], K. Ryu[6]

[1]*NASA Goddard Space Flight Center, Greenbelt, MD, USA*
[2]*University of Maryland Baltimore County, Center for Research and Exploration in Space Science and Technology, MD, USA*
[3]*Science Systems and Applications Inc., Lanham, MD, USA*
[4]*Sigma Space Corporation, Lanham, MD, USA*
[5]*ADNET Systems, Inc., Bethesda, Maryland, USA*
[6]*MIT Lincoln Laboratory MIT, 244 Wood Street, Lexington, MA USA*

We report on the development of multi-absorber transition edge sensors (TESs), referred to as 'hydras'. A hydra consists of multiple x-ray absorbers each with a different thermal conductance to a TES. Position information is encoded in the pulse shape. With some trade-off in performance, hydras enable very large format arrays without the prohibitive increase in bias and read-out components associated with arrays of individual TESs. Hydras are under development for the next generation of space telescope such as Lynx. Lynx is a NASA concept under study that will combine a < 1" angular resolution optic with 100,000-pixel microcalorimeter array with energy resolution of $\Delta E_{FWHM}$ ~ 3 eV in the soft x-ray energy range. We present first results from hydras with 25-pixels for Lynx. Designs with absorbers on a 25 µm and 50 µm pitch are studied. Arrays incorporate, for the first time, microstrip buried wiring layers of suitable pitch and density required to readout a full-scale Lynx array. The resolution from the coadded energy histogram including all 25-pixels was $\Delta E_{FWHM}$ = 1.66±0.02 eV and 3.34±0.06 eV at an energy of 1.5 keV for the 25 µm and 50 µm absorber designs respectively. Position discrimination is demonstrated from parameterization of the rise-time.

## 1 Introduction

X-ray spectroscopy is a powerful tool for probing the physical nature of matter in the universe at temperatures > $10^{5-6}$ K. Lynx is a NASA flagship mission concept currently under study that will provide answers into key scientific themes related to 1) the dawn of black-holes, 2) galaxy formation and evolution, and 3) stellar evolution and ecosystems [1]. The Lynx X-ray Microcalorimeter (LXM) is one of the main instruments, which will combine a sub-arcsecond x-ray optic, with 100,000-pixel microcalorimeter array (with $\Delta E_{FWHM}$ < 3 eV), and will provide exquisite imaging and spectroscopy capabilities.

The desire to implement a detector with 100,000 imaging elements introduces significant technical challenges. For example, if this were an array of individual TESs, the extremely high density of wiring within the array would be impractical, and the physical space required to incorporate the bias circuit and readout components would be prohibitively large. We are pursuing position-sensitive detectors as a practical approach to enabling 100,000-pixel arrays (which comes with some trade with resolution). Our 'hydras' consist of a single transition-edge sensor (TES) connected to multiple x-ray absorbers [2,3,4]. The absorbers are connected to the TES by a thermal 'link'. The conductance of each link is tuned to give a different thermal time constant between each absorber and the TES. Each pixel has a different characteristic pulse shape and enables position discrimination. Even with up to 25-pixels per hydra, the high-density, sub-µm wide traces required to connect all pixels within the array is challenging. We have developed a process to combine planarized multi-stack buried wiring layers from Massachusetts Institute of Technology Lincoln Laboratory (MIT/LL) with the NASA TES hydra. By combining 25-pixel hydras that incorporate buried wiring, with state-of-the-art microwave multiplexing [5], it becomes practical to realize a 100,000-pixel instrument within the engineering constraints of a satellite. We have previously reported on prototype hydra designs with up to 20 pixels per TES [4]. In this paper we report on the design and performance of the first 25-pixel hydras incorporating buried wiring, which are being specifically optimized for LXM.

* email: stephen.j.smith@nasa.gov
  Tel. : 301-286-3719

## 2 Hydra Designs for LXM

The current baseline LXM configuration [6] consists of a main array (MA) of 3456 hydras, each with 25 individual pixels on a 50 µm pitch (86,400 total pixels). The energy resolution goal is $\Delta E_{FWHM}$ ~ 3 eV at energies $E$ < 7 keV. The MA will provide a field-of-view of 5' where each pixel provides an angular resolution of 1". The central 1' region of the array will consist of 25-pixel hydras with each pixel on a 25 µm pitch (12,800 total pixels). This, enhanced main array (EMA), will provide 0.5" imaging with $\Delta E_{FWHM}$ ~ 2 eV.

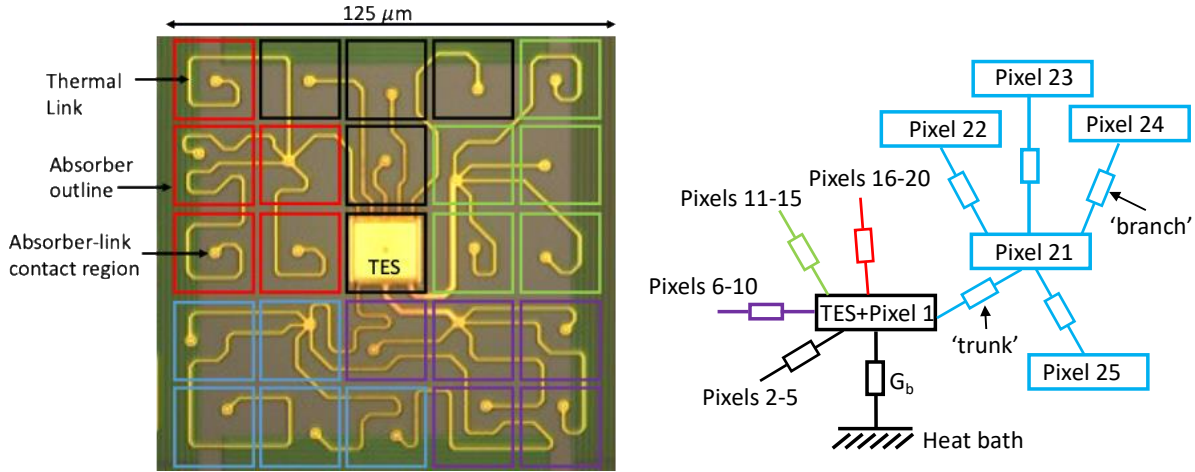

**Fig. 1** Left, photograph of a 25-pixel hydra with 25 µm pitch absorbers for the LXM EMA. This image was before absorbers have been deposited to show the hierarchical link layout. The outline of where the absorbers are later deposited are indicated. Each group of 5 pixels are indicated in a different color. Right, simplified thermal model of hydra 'tree' design.

We have fabricated a series of test arrays that include both the 250 µm pitch hydras and 125 µm pitch hydras in the same array. The TES is a 20×25 µm Mo/Au bilayer with normal state resistance of $R_n$ ~ 20 mΩ. The absorbers are 3.2 µm thick electroplated Au and are cantilevered ~ 4 µm above the substrate, supported by pillar shaped stems that make contact to the thermal links. The gaps between the pixels are 1.6 μm at the base of the absorbers and 5.2 μm at the top. The thermal links between the absorbers and the TES are $d$ = 310 nm thick evaporated Au and are either $w$ = 1 or 2 μm wide. The thermal conductance of a link at temperature T with length L is calculated from the Wiedemann-Franz relation $G$ = 24.5 $T$ ($d$ $w$) / ($\rho$ $L$) nW/K, where the resistivity of the metal link is $\rho$ = 0.67 μΩ cm. Thus, $L$ and $w$ are adjusted to tune the thermal conductance of each link. We use the same finite-element-modeling approach discussed in ref[3] to design the links, calculate the pulse shapes and the noise properties of the hydra. In [4] we showed a hierarchical 'tree' layout of the internal links that uses a series of 'trunks' and 'branches' to connect the pixels. This makes the design easier to layout when there are a large number of pixels to connect to the TES. In this design, instead of connecting every pixel directly to the TES, groups of pixels are connected together by link 'branches', and these groups then connect to the TES via link 'trunks' from 1 of the pixels in the group. For these designs, pixels are arranged into 5 groups, each with 5 pixels. Fig. 1 (left) shows the link layout for an EMA hydra before the absorbers have been deposited. Fig. 1 (right) shows a simplified thermal model. The full trunk and branch structure are only shown for pixels 21-25, which constitute Group 5. Fig. 2 (left) shows a photograph of part of a test array including both 125 μm and 250 μm hydra designs. Fig. 2 (right) shows a photo and cross-sectional schematic of how the TES connects to the buried microstrip Nb wiring layers using vias through the 200 nm thick $SiO_2$ layers. The Nb traces are 200 nm thick and 500 nm wide. These are of suitable geometry to be able to wire a full sized LXM array. We have previously used a buried Cu layer in NASA developed substrates to improve heat-sinking and reduce thermal-crosstalk between pixels in the array [7]. We are currently developing a process to incorporate heat-sink layers into the MIT/LL substrates; however, this is not included on the arrays discussed here, thus these devices are susceptible to additional noise from thermal cross-talk events.

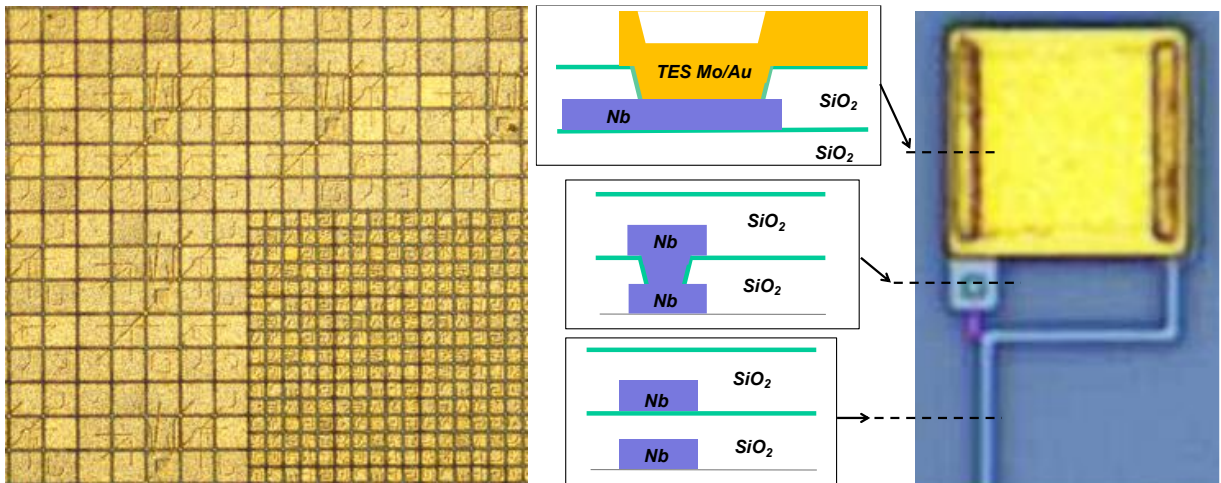

**Fig. 2** Left, photograph of part of the LXM test array including 125 μm pitch EMA hydras in the central region of the array surrounded by a region of 250 μm pitch MA hydras. Right, photograph with cross-sectional schematic of buried wiring layer connections to the Mo/Au TES.

## 2 Results

These devices were operated at 2.5% $R_n$ with a bath temperature of $T_b$ = 47 mK. The temperature of the TES under bias is $T_0 \sim$ 50 mK. The total heat-capacity is $C(T_0) \sim$ 0.50 pJ/K for the hydras with 50 μm pitch absorbers and $C(T_0) \sim$ 0.11 pJ/K for the 25 μm pitch design. In both cases this is dominated by the Au absorbers. The thermal conductance to the heat bath is $G_b(T_0) \sim$ 570 pW/K and $G_b(T_0) \sim$ 340 pW/K for the 50 μm and 25 μm designs respectively. Fig. 3 (left) shows the average measured pulse-shapes for a 25 μm design at an energy of 1.5 keV (Al-Kα x-rays). The inset shows a zoom-in of the first 40 μs. Because of the hierarchical layout of the link structure the pulse shapes have a more complicated pre-equilibration signal. Thus, the rising-edge of the pulse cannot be uniquely characterized by a single rise-time calculation. We are currently studying various techniques to extract position information from the pre-equilibration signal that are robust to the effects of pulse shape non-linearity with energy. Here we use a simple approach to parameterize the rising-edge of the pulse shapes using two metrics that extracts a fast and slow component to the rise-time. Fig. 3 (right) shows a scatter plot of the rise-time calculated from the 10-50% point of the pulse-height, versus the rise-time calculated from the 20-80% point after smoothing the pulse to suppress the faster component to the rise-time. This was sufficient to identify the 25 discrete populations of Al-Kα x-rays corresponding to each pixel in the hydra. This is not however a rigorously optimized algorithm and the broad-band position resolution is the subject of continuing studies.

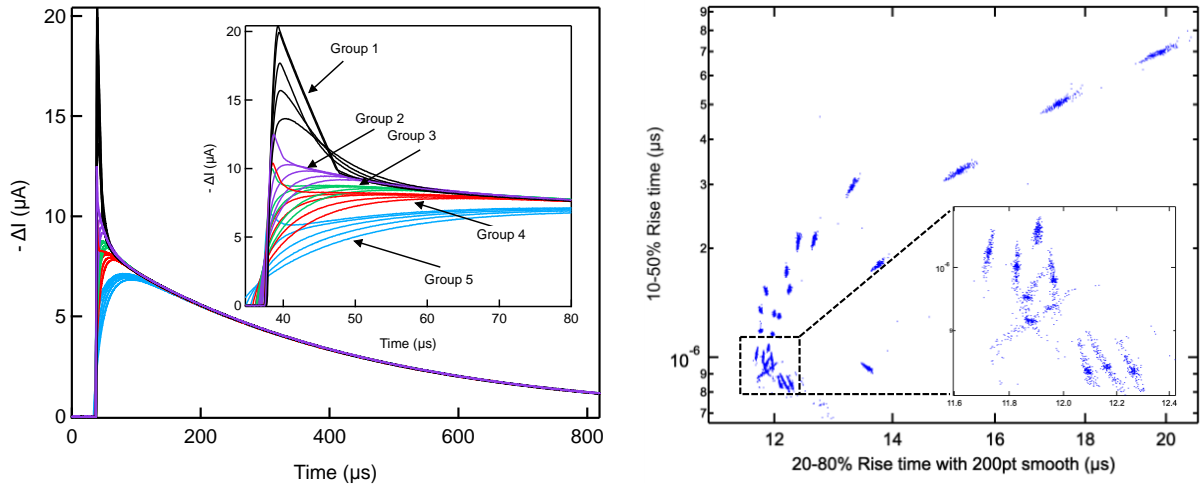

**Fig. 3** Left, average Al-Kα pulse shapes for an EMA hydra. The inset shows a zoom-in of the first 40 μs, with the different groups of pixels identified. Right, rise-time scatter plot for EMA pixel showing 25 separate regions that correspond to x-ray absorption in each of the 25-pixels.

Fig. 4 (left) shows the coadded energy histogram for all 25 pixels of an EMA hydra measured using a fluoresced Al-Kα x-ray target. The fitted resolution was $\Delta E_{FWHM}$ = 1.66±0.02 eV. This is consistent with the average of all pixels fitted individually, $\langle \Delta E_{FWHM} \rangle$ = 1.68±0.13 eV. Shown in Fig. 4 (right) is $\Delta E_{FWHM}$ from the individual energy histograms versus pixel number, illustrating a high degree of resolution uniformity. Shown for comparison is the resolution calculated from the integration over frequency of the noise-equivalent-power, $NEP(f)$ (where $NEP(f)$ is determined from measurements of the average pulse shape and noise spectral density). As expected, the integrated *NEP(f)* degrades slightly with increasing pixel number as the bandwidth of the pulse is reduced by the low-pass filtering from the links. By comparing the integrated *NEP(f)* calculated using noise measured with and without x-rays illuminating the array (shown on Fig. 4 (right)), we estimate that thermal cross-talk noise degrades the resolution by about 20% at the measurement count-rate of ~ 1.5 cps. Thus, it is reasonable to assume that for future designs with optimized heatsinking layers, improved $\Delta E_{FWHM}$ may be achievable. Because the resolution degradation due to cross-talk noise is proportional to *E*, we have not taken high statistics measurements at Mn-Kα (5.9 keV). However, low statistic measurements showed the average of the integrated *NEP(f)*, without the thermal cross-talk was 1.55 eV. This compares to 1.23 eV at Al-Kα and suggests that once the heat-sinking has been incorporated, $\Delta E_{FWHM}$ should be close to meeting the 2 eV design goal for energies up to 7 keV.

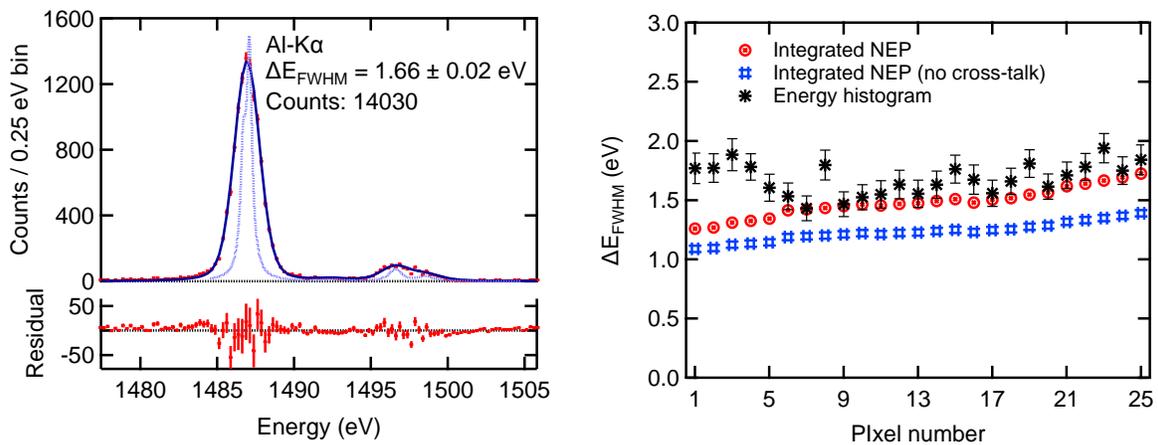

**Fig. 4** Left, coadded Al-Kα energy histogram for all 25 pixels of an EMA hydra. The best fit gives $\langle \Delta E_{FWHM} \rangle$ = 1.66±0.02 eV. Right, $\Delta E_{FWHM}$ versus pixel, including that from individually fitted energy histograms (black stars), and the predicted resolution from the integrated *NEP(f)*, measured both with (red circles) and without thermal cross-talk noise (blue boxes).

Similar preliminary tests have been carried out on a MA hydra with 50 μm pitch absorbers. These larger hydras have ~ 5-times the heat capacity of the EMA equivalents and the resolution should be approximately 2.5-times worse (since $\Delta E_{FWHM} \propto \sqrt{C}$). Thus, in the absence of any excess cross-talk noise we expect the detector performance to be around 3 eV. We measured $\Delta E_{FWHM}$ = 3.34±0.02 eV from the coadded energy spectrum for all 25 pixels for Al-Kα x-rays (Fig. 5). In this measurement the cross-talk noise is estimated to degrade the average resolution by ~ 20%, suggesting that ~ 3 eV could be achievable for a fully optimized device. The average measured pulse shapes are shown in Fig. 5 (right).

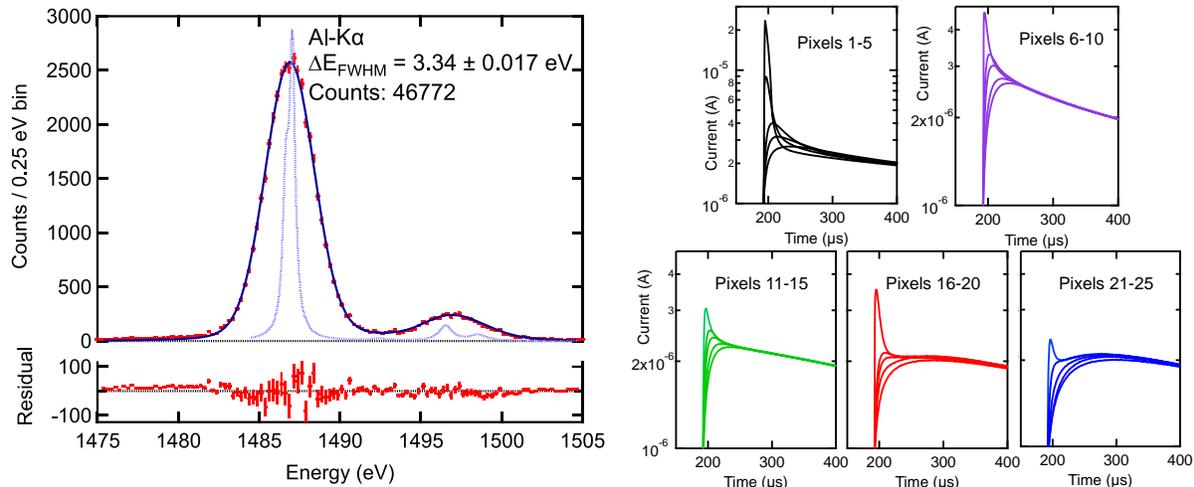

**Fig. 5** Left, coadded Al-K$\alpha$ energy histogram for all 25 pixels of an MA hydra. The best fit gives $\langle \Delta E_{FWHM} \rangle$ = 3.34±0.02 eV and the average of the individually fitted histograms was $\langle \Delta E_{FWHM} \rangle$ = 3.37±0.31 eV. Right, average measured pulse shapes for each of the 5 groups of pixels.

## 3 Conclusion

We have demonstrated the first results from 25-pixel hydras that include buried wiring layers. The initial designs are of suitable pixel and wiring pitch for a full sized LXM array and demonstrate the feasibility of making a 100,000-pixel instrument. Despite the lack of metallic heat-sinking layers these pixels demonstrated excellent energy resolution and the ability to distinguish all 25-pixels for hydras with 25 µm and 50 µm pitch absorbers for 1.5 keV x-rays. The broad-band (0.3-7 keV) energy resolution and position discrimination is the subject of ongoing testing. These measurements were made with low circuit inductance of ~ 20 nH and the rise-times were very fast. In order to multiplex many pixels, the rise-time must be slowed using higher circuit inductance. The impact of higher circuit inductance on the position resolution needs to be explored and will be the subject of future studies.